\documentclass{article}
\usepackage{latexsym,amsthm,amssymb}
\usepackage{mathdots,stmaryrd,ellipsis,multirow}
\usepackage{graphicx}
\usepackage{hyperref}

\newcommand{\St}{\ensuremath{\mathcal {S}}}
\newcommand{\In}{\ensuremath{\mathcal {I}}}
\newcommand{\Ot}{\ensuremath{\mathcal {O}}}
\newcommand{\U}{\ensuremath{\mathcal {F}}}
\newcommand{\K}{\ensuremath{\mathbf{K}}}
\newcommand{\evaluation}[2][]{\ensuremath{\llbracket #2\rrbracket_{#1}}}
\newcommand{\val}[2]{\evaluation[#2]{#1}}
\newcommand{\Dom}{\mathrm{Dom}~}
\renewcommand{\emptyset}{\varnothing}
\newcommand{\If}{\ensuremath{\mbox{\bf if}\;}}
\newcommand{\Then}{\;\mbox{\bf then}\;}
\newcommand{\True}{\textsf{true}}
\newcommand{\False}{\textsf{false}}
\newcommand{\Undef}{\textsf{undef}}
\newcommand{\Case}{\ensuremath{\textbf{case}\;}}
\newcommand{\Of}{\;\textbf{of}\;}
\newcommand{\When}{\textbf{when}\;}
\newcommand{\st}{\,|\,}
\let\Gam=\Gamma
\let\Del=\Delta
\renewcommand{\Gamma}{{\rm \Gam}}
\renewcommand{\Delta}{{\rm \Del}}
\newcommand{\restrict}{\!\upharpoonright\!}
\newtheorem{postulate}{Postulate}
\renewcommand{\thepostulate}{\Roman{postulate}}

\newtheorem{theorem}{Theorem}
\newtheorem{corollary}[theorem]{Corollary}

\newtheorem{lemma}[theorem]{Lemma}
\theoremstyle{definition}
\newtheorem{definition}[theorem]{Definition}
\newtheoremstyle{note}{3pt}{3pt}{}{}{\bfseries}{.}{.5em}{}
\newtheorem{remark}[theorem]{Remark}
\newtheorem{example}[theorem]{Example}
\renewenvironment{cases}{\left\{\begin{array}{ll}}{\end{array}\right.}
\newenvironment{pf}{\begin{proof}}{\end{proof}}
\newcommand{\hole}{\mbox{\Large$\bullet$}}

\title{Exact Exploration}
\author{Andreas Blass%
\thanks{Math Dept,  University of Michigan,  Ann Arbor, MI, USA.
Partially supported by NSF grant DMS-0653696.
Part of the work was performed at Microsoft Research.}
\and
Nachum Dershowitz%
\thanks{School of Computer Science, Tel Aviv University, Israel.
Partially supported by the Israel Science Foundation grant 250/05.
Part of the work was performed at Microsoft Research.}
\and
Yuri Gurevich%
\thanks{Microsoft Research, Redmond, WA\ 98052, USA.}}
\date{July 2009
\footnote{(A 2024 note:)
An abbreviated version of this paper was published in  CSL 2010, 19th EACSL Annual Conference on Computer Science Logic, Springer LNCS 6247, 2010, 140-154.}
}
\begin{document}
\maketitle
%\begin{ignore}
\begin{quote}\raggedleft\small%\singlespace
%It is only at the end of the last century that the somewhat vague principle of the economy of action or effort in physical actions \dots \\
%has culminated in the hands of Lagrange in his magnificent mathematical generalisation of the dynamical laws of material \\ systems.
Before the date of this [Lagrange's] concise and\\ 
all-embracing formulation  of the laws of dynamics\\ 
there was not available any engine of sufficient \\
power and generality to allow of a\\
thorough and exact exploration\\ of the properties of an ultimate medium,
of which\\ the mechanism and mode of action\\ are almost wholly concealed from view.
\\[1ex]

---\textit{Nature} (11 January 1894)
\end{quote}
%\end{ignore}

\begin{abstract}
Recent analysis of classical algorithms resulted in their axiomatization
as transition systems satisfying some simple postulates,
and in the formulation of the Abstract State Machine Theorem, %\cite{Gurevich00}
which assures us that any classical algorithm can be emulated step-by-step by
a most general model of computation, called an ``abstract state machine''.
We refine that analysis to take details of intra-step
behavior into account, and show that there is in fact an abstract state machine
that not only has the same state transitions as does a given algorithm but also
performs the exact same tests on states
when determining how to proceed to the next state.
This enhancement allows the inclusion---within the abstract-state-machine framework---of
algorithms
whose states only have partially-defined equality, or employ other native partial functions,
as is the case, for instance, with inversion of a matrix of computable reals.
\end{abstract}

%\category{F.1.0}{Theory of Computation}{COMPUTATION BY ABSTRACT DEVICES}[General]
%\terms Sequential Algorithms, Axiomatization, Programming, Emulation, Theory
%\keywords

\newpage
\setcounter{tocdepth}{2}
\tableofcontents

\newpage
\section{Introduction}

\textit{Abstract state machines (ASMs)}
\cite{Gurevich94b} constitute a most general model of (sequential) computation,
which can operate on any level of abstraction of data structures and native operations.
By virtue of the Abstract State Machine Theorem of \cite{Gurevich00} (henceforth the ``original study''),
any algorithm that satisfies three ``Sequential Postulates''
can be step-by-step emulated by an ASM.
These postulates formalize the following intuitions: (I) we are talking about deterministic state-transition systems;
(II) the information in states suffices to determine future transitions and may be captured by logical structures that respect isomorphisms;
and (III) transitions are governed by the values of a finite and input-independent set of (variable-free) terms.

A careful analysis of the notion of algorithm in the original study,
as well as an examination of the intent of the founders of the field of computability in \cite{CT},
have demonstrated that the Sequential Postulates are in fact
true of all ordinary, sequential algorithms, the (only) kind envisioned
by the pioneers of the field.

Our goal in the current endeavor is to refine the previous analysis,
axiomatization and theorem to take into account the precise set of
locations in each state that are accessed or examined by the algorithm.
This may be critical when dealing, for example, with objects like computable reals,
for which inequality may be only partially computable, so cannot be used indiscriminately.
The proposed refinement
should contribute to the belief that ASMs are a universal model of sequential computation
in the very strong sense of precise emulation.

In Section~\ref{sec:back}, we recapitulate some of the analysis of the classical notion of algorithm from the above-cited works.

In an effort to be self-contained,
we briefly review the three original Sequential Postulates, ASM programs, and the ASM Theorem
in Sections~\ref{sec:ax} and~\ref{sec:asm}, and consider how they ought to be modified.

The refined third axiom, restricting exploration, is developed in Section~\ref{sec:exp}
and compared with the original version of bounded exploration.
The main result, that
for every algorithm there is a behaviorally equivalent ASM that
explores the exact same set of locations as does the given algorithm,
is shown by construction in Section~\ref{sec:le},
which includes a definition of what it means to be ``behaviorally equivalent''
when we are also interested in precisely which locations are explored.

Sections \ref{sec:part} and \ref{sec:case} extend the analysis to allow for failed exploration, partial functions, and multivalued tests.
Variants and consequences of the refined exploration axiom are touched on in Section~\ref{sec:vary},
followed by a brief discussion
of the implications of this work for the Church-Turing Thesis.

But first, we explain the importance of this foundational study for the understanding of algorithms and computation.

\section{Significance}

The significance of the ASM Postulates lies in their comprehensiveness.
They formalize which features exactly characterize a sequential algorithm in its most
abstract and generic manifestation.
All models of effective, sequential computation satisfy the postulates, as
do idealized algorithms for computing with real numbers, or for geometric constructions
with compass and straightedge.
See \cite{Reisig04} for some examples.

Abstract state machines are a computational model that
is not wedded to any particular data representation, in the way, say,
that Turing machines manipulate strings using a small set of tape operations.
The ASM Theorem of the original study proves that ASMs can express
any and all algorithms satisfying the premises captured by the postulates.
%Are ``the premises captured by the postulates'' intended to be different from just ``the postulates''? 
For any such algorithm, there is an ASM program that describes precisely the
same state-transition function as does the algorithm.
In this sense, ASMs subsume all other computational models.

There are at least three important reasons for delving into the issue of precise emulation:
universality of ASMs for describing algorithms,
fidelity of ASMs to the inner workings of algorithms,
and parsimony of the description of an emulating ASM.

\paragraph{Universality.}
The states of standard ASMs always come with an equality relation between all base-set elements.
Furthermore, operations in states are always total, with
partiality represented by explicit values for ``undefined''.
With the refinements developed here,
one can naturally model all varieties of sequential algorithms,
with total or partial operations,
and even with only partially defined equality
(which might vary from initial state to initial state, depending on the inputs).

Additionally,
one can now model system-wide failure authentically: if any part of the program
attempts a zero division, for example,
the computation as a whole gets stuck in an unresponsive state,
what we will call a ``black hole''.
See Section~\ref{sec:part}.

Furthermore,
the results described herein serve to bolster the belief that the Sequential Postulates succeeded in
capturing all sequential algorithms---as claimed in the original study---%
regardless of which model of computation they
may be expressed in, by showing that the postulates also faithfully cover algorithms that
employ native operations that are only partially defined.
%Returning to our motivating example:
One can, for example, work with genuine (infinite-precision computable) reals, in symbolic form,
for which testing for zero is undecidable.

\paragraph{Fidelity.}
The ASM Theorem presupposes the availability of an equality test, which is used in the guards of commands in the emulating ASM.
This paper sheds light on how to emulate algorithms even when
only limited equality between values is actually available.

For example, a Gaussian elimination program would test that a pivot element $p$ is non-zero
before dividing an array element $a[i,j]$ by it.
Since the program would include a statement involving the expression $a[i,j]/p$,
the emulating ASM, as produced by the proof of the ASM Theorem,
would include that expression in conditions that are always evaluated,
regardless of the value of $p$, which is undesirable.
It is clear that there is an ASM that first tests $p$ and only when $p\neq 0$ needs to look
at $a[i,j]/p$.
And indeed, the
emulating ASM constructed in Section~\ref{sec:le} works that way;
see Section~\ref{sec:eq}.

Similarly, an algorithm for inverting a matrix of (computable) real (or complex) numbers, by
first computing its adjugate (classical adjoint) and then dividing through by
the value of the determinant,
might be expressed in terms of
arithmetic operations on the reals, without ever testing their equality or disequality.
%(The computable reals form a field.)
Likewise, states might only
partially interpret various other function symbols, besides equality, like division.
In particular, in the adjugate matrix inversion method, only if the determinant
is zero ought the result of inversion be undefined.
See Section~\ref{sec:op}.

\paragraph{Parsimony.}
Querying a state about the values of its locations
may be time consuming and expensive. Why? Because states are abstractions
of the data that are potentially available,
whereas,
in reality, an implementation may need to investigate its environment to actually obtain those values,
or may need to invest great effort in reconstructing them.
In that case, one would not want the emulating ASM to explore parts of the state that the original algorithm does not.
But the ``normal-form'' ASM that is constructed as part of the proof of the ASM Theorem
%in \cite{Gurevich00}
performs many tests that might not all be necessary
for the determination of the next state.

For example, an algorithm for removing duplicates from a file system
may need to sometimes test equality of gigantic files, but would first check to
see that their recorded sizes are the same.
The normal-form ASM, however, would always check both size and content,
despite the tremendous overhead.
On the other hand, the
emulating ASM constructed in Section~\ref{sec:le} avoids such tests.
By eliminating unnecessary tests, the emulating ASM program is often simpler and shorter---%
with no need for human ingenuity to improve the normal-form ASM obtained directly from the theorem.%
\footnote{Cf.\ the discussion of algorithm equivalence in \cite{PITS},
and the illustration therein of two ASMs with the same state-for-state behavior.}
%one of which always explores everything, while the other has a tree-like exploration.}

\section{Background}\label{sec:back}

\begin{figure}
\[
\renewcommand\arraystretch{1.4}
\begin{array}{l}
[~\If j \neq n \Then\\
\quad [~\If F(i)>F(j) \Then [F(i) := F(j)  \,\|\, F(j) := F(i) ]\\
~~~~ \|\; j := j+1 ~]\\
 \;\|\; \If j = n \wedge i+1 \neq n \Then [i := i+1 \;\|\; j := i+2~]~]
\end{array}
\]
\caption{An abstract-state-machine program for sorting.}\label{fig}
\vspace*{6pt}\hrule
\end{figure}

The ASM Postulates assert that a classical algorithm is a state-transition system operating over first-order structures in a way that is invariant under isomorphisms.
Thus a state $X$ interprets each function symbol $f$ as an operation over its \emph{base set} (\emph{underlying set}, or \emph{domain}) $\Dom X$, and in that way gives
a meaning $\val{t}{X}\in\Dom X$ to every term $t$.
(Whenever we speak of a ``term'', we will mean a \emph{ground} term---sans variables.)

An algorithm is a prescription for updating states, that is, for changing some of the interpretations given to symbols.
The essential idea is that there is a fixed finite set of terms that refer (possibly indirectly) to locations within a state
and which suffice to determine how the state changes during a transition.
The actions taken by a transition are describable in terms of updates
of the form $f(\bar a)\mapsto b$ meaning that $b$ is the new interpretation to be given by the state
to the function symbol $f$ for values $\bar a$.

For example, the state of a sorting algorithm
may have integers in its base set, along with some static arithmetic and logical operations.
Fixed nullary functions $0$ and $n$ (programming ``constants'') can serve as bounds of an array $F$,
where $F$ is a unary function; in addition
varying nullary functions $i$ and $j$ (programming ``variables'') can be used as array indices.
Initially $i=0$ and $j=1$,
and the algorithm proceeds by modifying the values of $i$ and $j$ as well as of locations
$F(0),\dots,F(n-1)$, by referring to terms $F(i)$ and $F(j)$.
See Figure~\ref{fig} and Example~\ref{ex:sort} in Section~\ref{sec:asm} below.

We adopt most of the analysis of classical
algorithms in previous work on ASMs.
In particular, we observe the following points:
\begin{itemize}
\item A state (like the ``instantaneous description'' of a Turing machine computation) contains \emph{all} the relevant information, besides the algorithm itself, needed to determine the next steps.
\item {The values of ``programming variables'', in and of themselves, are meaningless to an
algorithm, which is implementation independent.  It is the relationships between values that matter to the algorithm.
Accordingly, a state stores values in its locations in some internal format and provides
the algorithm with access to those values
in the form of concrete answers to queries about their relationships.
\item An algorithm must access the state and sometimes change values stored therein.
We speak about this interaction on the precise level of the abstraction of the algorithm, independent of any specific implementation of states.
In this sense,
the interface between algorithm and state is ``public'' and ``objective''.}
\item First-order structures suffice to model all salient features of states.
The only means an algorithm has at its disposal for determining relations between values stored in a state is via terms.
\item Algorithms are expressed by means of finite texts, making reference to the values of only finitely many terms.
\end{itemize}

In contrast with the original study, we will not necessarily presume here that states are always endowed with the equality relation for all pairs of elements of their base set, nor that states contain values for all function symbols applied to all tuples of elements (of the appropriate length) of its base set.
These considerations will be taken up in Section~\ref{sec:part}.

A classical ASM typically models partial functions by using a special value, \Undef, denoting
that the argument is outside the function's domain of definition, and arranging
that all operations be strict, so any term involving a subterm that is undefined is also undefined.
The state of a classical ASMs would return \True\ when asked to evaluate
an expression $a[i,j]/p = \Undef$, when $p=0$, and so it can be programmed to work properly,
despite the partiality of division.
It is usually an easy matter in applications to include ``weak equality'' in states,
under which testing for equality of a defined value with an undefined value always yields \Undef.
But it is, of course, better not to count on a proper implementation,
and to have the model itself enforce faithfulness to the notion of truly partial functions,
which return no value at all and whose domain of definition may be undecidable.
We return to this issue in Section~\ref{sec:ex}.

We deal here only with the ``classical'' type of algorithms, that is to say, with the ``small-step'' (meaning, only bounded parallelism) ``sequential-time'' (deterministic, no intra-step interaction with the
outside world) case,
called ``sequential algorithms''.
In \cite{OrdI,OrdII,OrdIII,General1,General2},
the analysis of sequential algorithms was extended to the case when the algorithm may interact with the outside environment during a step. We do not consider such intra-step interaction with the outside environment here.
But there is also an internal interaction between (the executor of) the algorithm and the state of the algorithm.
In fact, different implementations of the algorithm may have different implementations (``representations'') of the state.
Though we abstract from implementation details, the need for an algorithm to interact with the state remains.
This internal interaction is much simpler than the intra-step interaction with the external environment analyzed in the cited works, yet goes beyond that of the original study.
Hence the need for this study.
See Section~\ref{sec:model}.
%More modern types of algorithms, involving as they do interaction with the environment and
%unbounded parallelism, require additional analysis.

\subsection*{Related Work}

We are aiming for a model of computation that can faithfully support 
algorithms for which basic operations may have varying costs involved,
and/or for which their domains of applicability may be unknown or uncomputable. 
\looseness=-1
The foundation built here provides an operational semantics for
programming with objects like computable reals, represented by partial algebras.
See, for example, \cite{WeihrauchAnalysis}.
There are many implementations of arithmetic with infinite-precision reals, including
xrc in C (see \url{keithbriggs.info/xrc.html} and other links there)
and a Lisp package (\url{www.haible.de/bruno/MichaelStoll/reals.html}).
See also \cite{survey,Real}.
And there are optical models with some arithmetic but no equality
\cite{Optical}.
%\nb{geometry example?}

%On the practical side,
ASMs have been used to model all manner of
programming applications, systems, and languages, each
on the precise intended level of abstraction.
See \cite{Boerger02a} and the site \url{www.eecs.umich.edu/gasm}.
AsmL \cite{AsmL},
an executable specification language based on the ASM framework,
has been used in industry, in particular for the behavioral specification of interfaces \cite{BarSch01b}.
ASMs have been used in \cite{Nowack} to model the BSS model of computation with real numbers  \cite{BSS}.
The work herein provides theoretical justification
for the applicability of the ASM paradigm also to domains with partial functions,
and allows for the specification of programming languages with such features,
by means of interpreters expressed as ASMs.

\section{Axiomatization of Algorithms}\label{sec:ax}

In this section, we briefly recount the original postulates % of \cite{Gurevich00} % and \cite{OrdI}
regarding algorithmic behavior,
taking the output of algorithms explicitly into account.
With an eye on the considerations outlined in the previous section,
we refine those postulates.

\subsection{Sequential Time}

To begin with,
algorithms are deterministic state-transition systems.

\begin{postulate}[Sequential Time]\label{P1}
An algorithm determines the following:
\begin{enumerate}
\item A nonempty set\/\footnote{Or class.}
$\St$ of \emph{states}, a nonempty subset\/\footnotemark[\value{footnote}] $\In \subseteq \St$ of \emph{initial states}, and a subset\/\footnotemark[\value{footnote}] $\Ot \subseteq \St$ of \emph{terminal states}.
\item A \emph{next state} transition function $\tau:\St\setminus\Ot\rightarrow\St$.
\end{enumerate}
\end{postulate}
Alternatively, one may think of $\tau$ as a partial function $\tau:\St\rightharpoonup\St$,
which is only defined for non-terminal states.
So, we can express that $X\in\St$ has no next state by way of
$\tau(X)=\bot$.
Terminal states $\Ot$ can come in two varieties, \emph{successful} and \emph{failing}.

This postulate asserts that we are dealing with state-transition systems.%
\footnote{It is just like the Sequential Time Postulate of the original study,
except that this version insists that there are in fact some initial states (or else there would be no computations)
and also takes into consideration the possibility that an algorithm may halt---whether with success or with failure---without producing a next state.}

Having the transition $\tau$ depend only on the state
means that states must store all the information needed to determine subsequent behavior.
Prior history is unavailable to the algorithm unless stored in the current state.

Classical algorithms never leave room
for choices, nor do they involve any sort of interaction with the environment to determine
the next step.%
\footnote{Bounded nondeterminism is dealt with in \cite{GY}.}
Hence, we analyze only deterministic transition systems here.

\subsection{Abstract State}

States may be viewed as
first-order structures (or ``partial algebras'' in the sense of universal algebra).
Each state consists of a domain and interpretation for symbols.
All relevant information about a state
should be given explicitly in the state by means of its interpretation of the function and relation symbols
appearing in the vocabulary of the structure.
The specific details of the implementation of the data types used by the algorithm should not matter.
In this sense states are ``abstract''. This crucial consideration leads to the second postulate.

\begin{postulate}[Abstract State]\label{P2}
The states $\St$ of an algorithm are (first-order) structures,
possibly with partial operations,
over a finite vocabulary $\U$, such that the following hold:
\begin{enumerate}
\item If $X\in \St$ is a state of the algorithm, then any structure $Y$ isomorphic to $X$ is also a state in $\St$, and $Y$ is initial or terminal if $X$ is initial or terminal, respectively.
\item Transitions $\tau$ preserve the base set; that is, $\Dom\tau(X)=\Dom X$ for every non-terminal state $X\in \St\setminus\Ot$.
\item Transitions respect isomorphisms, so, if $\zeta:X\cong Y$ is an isomorphism of non-terminal states $X,Y\in \St\setminus\Ot$, then $\zeta: \tau(X)\cong\tau(Y)$.
\end{enumerate}
\end{postulate}

This postulate is justified by the same considerations as in the original study,
namely, the vast experience of mathematicians and scientists who have faithfully and transparently
presented every kind of static mathematical/scientific reality as a logical structure.
Closure under isomorphism ensures that the algorithm can
operate on the chosen level of abstraction
and remain oblivious of the internals of states.
So states are ``comprehensive'':
they incorporate all the relevant data (including any ``program counter'')
that, when coupled with the program, completely determine the future of a computation,
but the states' internal representation of the data is invisible and immaterial to the program.
Vocabularies are finite, since an algorithm must be describable in finite terms,
so can only refer explicitly to finitely many operations.%
\footnote{The only differences between this and the original Abstract State Postulate
 are 
 that operations may be partial and provision has been made for computations that explicitly halt in terminal states.
In the original study, a computation was viewed as ``completed''
when a state transitions to itself.
Most quotidian algorithms, however, halt explicitly in an observable terminal state,
a situation that should be distinguished from when algorithm gets ``stuck'' in a fixpoint loop.}

Since a state $X$ is a structure, it interprets function symbols in $\U$,
assigning a value $b$ from $\Dom X$ to the ``location'' $f(a_1,\dots,a_k)$ in $X$ for every $k$-ary symbol
$f\in\U$
and for those values $a_1,\dots,a_k$ in $\Dom X$
for which $f$ is defined.
In this way, $X$ assigns a value $\val{t}{X}$ in $\Dom X$ to terms $t$ over $\U$
(as long as all the symbols in $t$ are defined at the relevant points; see Section~\ref{sec:part}).

It is convenient to view each state as a collection of the graphs of its operations,
given in the form of a set
of location-value pairs, each written conventionally as % ``assignments'' of the form
$f(\bar a) \mapsto  b$, for $\bar a,b\in\Dom X$.
Define the \emph{update set} $\Delta(X)$ of state $X$ as the changed points, $\tau(X)\setminus X$.
When $X$ is a terminal state and $\tau(X)$ is undefined, then we will indicate that by setting
$\Delta(X)=\bot$ (that is, undefined).

An algorithm can make an explicit distinction between successful and failing
terminal states by storing particular values
in specific locations of the final state.

We will also need to handle the possibility that an algorithm ``hangs'',
waiting helplessly for a response from the state.
To distinguish between knowing that there is no next state, indicated by $\Delta(X)=\bot$,
and not knowing that there is none,
as in this case, we let $\Delta$ also take on a ``black hole'' value, $\hole$.
See Section~\ref{sec:part}.

The point is that $\Delta$ encapsulates the state-transition relation $\tau$ of an algorithm
by providing all the information necessary to update the interpretation given by the current state.
But to produce $\Delta(X)$, the algorithm needs to evaluate, with the help of {the information stored in} $X$, the values of some terms.
Later, we will use $\Gamma(X)$ to refer to the set of these ``exploration'' terms.
The next postulate will ensure that $\Delta$ has a finite representation and its updates can be performed
by means of only a finite amount of work.

\subsection{Bounded Exploration}

The original third postulate simply states that there is a fixed, finite set of ground (variable-free) terms that determines the behavior of the algorithm.

\begin{postulate}[Bounded Exploration]\label{BE}
An algorithm with states $\St$ over vocabulary $\U$
determines a finite set $T$ of \emph{critical terms}  over $\U$, such that
states that agree on the values of the terms in $T$ also share the same update sets.
That is,
\begin{equation}\label{T}
\mbox{if } X =_T Y \mbox{ then } \Delta(X)=\Delta(Y)\;,
\end{equation}
for any two states $X,Y\in\St$.
\end{postulate}
\noindent
Here,
$X =_T Y$, for a set of terms $T$, means that $\val t X = \val t Y$ for all $t\in T$.
We will express this by saying that structures $X$ and $Y$ \emph{agree} on the values of critical terms $T$.
In what follows, we will presume that the set $T$ of critical terms is closed under subterms.

%$X \equiv_T Y$, for a set of terms $T$, means that for each pair of terms $s,t\in T$,
%it's the case that
%$\val s X = \val t X$ iff $\val s Y = \val t Y$.
%In other words, the structures $X$ and $Y$ agree on the partition of $T$ into congruence classes.

The intuition is that an algorithm must base its actions on the
values contained at locations in the current state.
Unless all states undergo the same updates unconditionally, an algorithm must
explore one or more values at some accessible locations in the current state before determining how to proceed.
The only means that an algorithm has with which to reference locations is via terms,
since the values themselves are abstract entities.
If every referenced location has the same value in two states,
then the behavior of the algorithm must be the same for both of those states.
%The only aspect of terms that is implementation-independent is
%equality between the values given them by a state.
Subsequent actions may include---besides updates themselves---the act of exploring different locations.

\subsection{Classical Algorithms}

All classical algorithms satisfy the above postulates.
We formalize this observation in the following definition:
\begin{definition}[Classical Algorithm]
An algorithm
satisfying Postulates \ref{P1}, \ref{P2}, and \ref{BE}
will be called \emph{classical}.
\end{definition}
\noindent
In this sense, the traditional notion of algorithm is precisely captured by the three postulates.

In Section~\ref{sec:exp}, we will revise the third postulate,
since we are interested in the more refined set of explored terms $\Gamma(X)$,
rather than the full set $T$ of critical terms as in the above version.
What we want is a stronger, localized version of~(\ref{T}), namely:
\begin{equation}\label{eq}
\mbox{\it if } X=_{\Gamma(X)} Y \mbox{\it\ then } \Delta(X)=\Delta(Y)\;.
\end{equation}
We will actually need both aspects of behavior---exploration, as well as updates---to be fully
determined by the values of terms that are actually explored.
So, in addition, we will demand that
\begin{equation}\label{eq2}
\mbox{\it if } X=_{\Gamma(X)} Y \mbox{\it\ then } \Gamma(X)=\Gamma(Y)\;.
\end{equation}

\section{Abstract State Machines}\label{sec:asm}

Abstract State Machines (ASMs) are an all-powerful language for
classical algorithms.
{For convenience,
we employ a simple form of ASMs below.
(The reader should
bear in mind that
richer languages for ASMs are given in \cite{Gurevich94b} and are used in practice.)}

\subsection{ASM Programs}\label{sec:prog}

Programs are expressed in terms of some vocabulary, which---we may always assume---includes
symbols for the Boolean values (\True\ and \False), standard Boolean operations
($\neg$, $\wedge$, $\vee$), and equality ($=$).

\begin{definition}[ASM]
An \emph{ASM program} $P$ over a vocabulary $\U$ is a finite text, taking one of the following forms:
\begin{itemize}
\item An \emph{assignment} statement $f(s_1,\dots,s_n) := t$, where $f\in \U$ is a function symbol of arity $n$,
$n\geq 0$, and the $s_i$ and $t$ are ground terms over $\U$.
\item A \emph{parallel} statement $[P_1 \;\|\; \cdots \;\|\; P_n]$ ($n\geq 0$), where each of the $P_i$ is an ASM program over $\U$.
(If $n=0$, this is ``do nothing'' or ``skip''.)
\item {A \emph{conditional} statement $\If C \Then P$, where $C$ is a Boolean condition over $\U$, and $P$ is
an ASM program over $\U$.}
\end{itemize}
\end{definition}
\noindent
The semantics of these ASM statements are as expected, and are formalized below.
The program, as such, defines a single step, which is repeated forever or until
there is no next state.
%\nb{Cf. Kleene and "While Theorem".}
%Can leave connectives out of \U but in conditions

\begin{example}\label{ex:sort}
Let $\U=\{1,2,+,>,=,F,n,i,j\}$ be the vocabulary of a sorting program.
By default, ASM programs also include symbols for $\True$, $\False$, and \Undef\ for ``undefined'',
and for the standard Boolean operations.%
\footnote{It is not absolutely necessary for the states themselves to harbor the
Boolean operations.  One could consider them, instead, to be part
of the programming language syntax only, used for forming conditionals.\label{Bool}}
Let all states interpret the symbols $1,2,+,>,=$, as well as the default symbols, as usual.
These are static; their
interpretation will never be changed by the program.
Let initial states have $n\geq 0$, $i=0$, $j=1$, some integer values for $F(0),\dots,F(n-1)$,
plus \Undef\ for all other points of $F$.
Figure~\ref{fig} displays a simplified selection-sort in this language,
where $j\neq n$ is short for $\neg(j=n)$.
This program rearranges $F$
so that  $F(0)\leq F(1) \leq \cdots \leq F(n-1)$ in the end.
It always terminates successfully, with $j=n=i+1$
and with the first $n$ elements of $F$ sorted.
\qed\end{example}

We point out that every such ASM program can be reformulated as a single parallel application
$[P_1\|\cdots\|P_n]$,
where each $P_i$ is a nested conditional assignment of the form
\[\If C_1 \Then \If C_2 \Then \cdots \Then f(s_1,\dots,s_n) := t\]
(or nothing, $[\,]$, in place of the assignment).
This is accomplished by repeatedly replacing
$\If C \Then [P_1\|\cdots\|P_n]$
with the semantically equivalent
$[\If C \Then P_1\|\cdots\| \,\If C \Then P_n]$.
For any state $X$, the exact same conditions are evaluated and assignments executed.

\begin{table}[t]
\[
\renewcommand\arraystretch{1.4}
\begin{array}{|l||c|c|}\hline
  & \mbox{States $X$ such that} & \mbox{Update set $\Delta(X)$}\\\hline\hline
0 & \val{j}{}=\val{n}{}=\val{i}{}+1 & \bot\\\hline
1 & \val{j}{}=\val{n}{}\neq\val{i}{}+1 & i\mapsto \val{i}{}+1 ,\;  j\mapsto \val{i}{}+2\\\hline
2 & \val{j}{}\neq\val{n}{},\;\val{F(i)}{}>\val{F(j)}{} & 
\begin{array}{c}
F(\val{i}{})\mapsto \val{F(j)}{},\;F(\val{j}{})\mapsto \val{F(i)}{}, \\  j\mapsto \val{j}{}+1
\end{array}
\\\hline
3 & \val{j}{}\neq\val{n}{},\;\val{F(i)}{}\leq\val{F(j)}{} & j\mapsto \val{j}{}+1\\\hline
\end{array}
\]
\caption{Update sets for the sorting example (the subscript in $\val{\cdot}X$ is omitted).}\label{tab:T}
\vspace*{3mm}\hrule
\end{table}

\subsection{Update Sets of ASMs}\label{sec:up}

Unlike algorithms, which are observed to
either change the value of a location in the current state, or not,
an ASM might ``update'' a location in a \emph{trivial} way, giving it the same value it already has.
Also, an ASM might designate two conflicting updates for the same location, in which case
the standard ASM semantics are to cause the run to fail.%
\footnote{An alternative semantics, namely, nondeterministic choice between values,
was also considered in \cite{Gurevich94b}.}

To take these additional possibilities into account,
a \emph{proposed} update set $\Delta^+_P(X)$ for an ASM $P$ may be defined in the following manner:%
\footnote{This notion of proposed updates $\Delta^+$ arose in \cite{OrdI},
where 
it was pointed out that,
when algorithms are distributed and more than one process may be vying
for access to a location,
trivial updates can cause an observable difference.}
\begin{equation}\label{eq:prop}
\renewcommand\arraystretch{1.3}
\begin{array}{rcl}
\Delta^+_{f(s_1,\dots,s_n) := t}(X) &=& \{f(\val{s_1}X,\dots,\val{s_n}X) \mapsto  \val tX\}\\[2pt]
\Delta^+_{[P_1 \;\|\; \cdots \;\|\; P_n]}(X) &=& \Delta^+_{P_1}(X)\cup\cdots\cup \Delta^+_{P_n}(X)\\
\Delta^+_{\mbox{\bf\small if} \;C\;\mbox{\bf \small then}\,P}(X) &=&
\begin{cases}
\Delta^+_P(X) & \mbox{if $X\models C$}\\
\emptyset & \mbox{otherwise}\;.
\end{cases}
\end{array}
\end{equation}
When the condition $C$ of a conditional statement does not evaluate to $\True$,
the statement does not contribute any updates.
When $\Delta^+_P(X)$ contains inconsistent updates,
$f(\bar a)\mapsto b$ and $f(\bar a)\mapsto b'$ with $b\neq b'$,
we set $\Delta^+_P(X)=\bot$, and say that
the ASM $P$ ``fails'' and provides no next state;
when $\Delta^+_P(X)=\emptyset$, it halts with success.
In either case, $X$ is a terminal state.
Otherwise, the updates are applied to $X$ to yield the next state,
by replacing the values of all locations in $X$ that are referred to in $\Delta^+_P(X)$.
So, if the latter
contains only trivial updates, $P$ will loop forever.

Let $\Delta^0(X)$ denote the set $\{f(\bar a)\mapsto  \val{f(\bar a)}{X} \st \bar a \in \Dom X \}$
of all possible trivial updates for state $X$.
Then, the update sets $\Delta(X)$ for the algorithm given by an ASM program $P$ can be derived from $\Delta^+$ as follows:
\[
\Delta(X) =
\begin{cases}
\bot & \mbox{if $\Delta^+(X)\in\{\emptyset,\bot\}$}\\
\Delta^+(X) \setminus \Delta^0(X) & \mbox{otherwise}\;.
\end{cases}
\]
(As long as no confusion will arise, we are dropping the subscript $P$.)
Let
\[
 \Delta^-(X) = \{f(\bar a)\mapsto \val{f(\bar a)}{X} \st f(\bar a) \mbox{ is updated in } \Delta(X)\}\
\]
be the set of location-value pairs of all locations slated to be changed.
The next state is the result
\[
\tau(X) = \left(X\setminus \Delta^-(X)\right) \cup\Delta(X)
\]
of applying those updates,
when $\Delta(X)\neq\bot$, and is undefined, otherwise.

\begin{example}
The update sets for the above sorting program are given in Table~\ref{tab:T}.
For example, if state $X$ is such that  $n=2$, $i=0$, $j=1$, $F(0)=1$, and $F(1)=0$,
then (per row 2)
$\Delta^+(X)=\{F(0)\mapsto 0, F(1)\mapsto 1, j\mapsto 2\}$.
For this $X$,
$\Delta(X)=\Delta^+(X)$, and
the next state $X'=\tau(X)$ has
$i=0$, $j=2$, $F(0)=0$ and $F(1)=1$.
After one more step (per row 1), in which $F$ is unchanged,
the algorithm reaches a successful terminal state, $X''=\tau(X')$, with $j=n=i+1=2$.
Then (by row 0), $\Delta^+(X'')=\emptyset$ and $\Delta(X'')=\bot$.
(This program never fails, as $\Delta^+$ never includes inconsistent updates.)
\qed\end{example}

\subsection{The ASM Theorem}

Abstract state machines clearly satisfy Postulates \ref{P1}--\ref{BE}.
%(\cite{Gurevich00}, with slight modifications along the lines of \cite[Part III]{Ords} to accommodate halting and failing computations).
ASMs define a state-transition function; they operate over abstract states;
and they depend critically on the finite set of terms appearing in the program.

\begin{example}
The critical terms for our sorting example (Figure~\ref{fig}) are all the terms in the program,
except for the left-hand sides of assignments, which contribute their proper subterms instead.
These are
$ j\neq n$, $(j = n) \wedge (i+1 \neq n)$, $F(i)>F(j)$, $i+2$, $j+1$,
and their subterms.
Only the values of these affect the computation.
\qed\end{example}

\begin{theorem}[ASM Theorem {\cite[Theorem 6.13]{Gurevich00}}]\label{thm:asm}
Every classical algorithm, satisfying Postulates \ref{P1}--\ref{BE},
has an equivalent ASM,
with the exact same states and state-transition function.
\end{theorem}

The proof of this theorem constructs an ASM that contains conditions involving equalities and disequalities between all the critical terms.  These conditions can be very large and complicated.
Theorem~\ref{thm} below is a refinement that avoids unnecessarily complicated conditions.

\begin{example}\label{ex7}
Given the above critical terms and
sort algorithm, the ASM constructed by the proof of the
ASM Theorem would include statements like
\[
\If (F(i)>F(j))=\True \wedge j = n \wedge i+1 \neq n \Then j := i+2 \;.
\]
This, despite the fact the first conjunct of the conditional is irrelevant when the other two hold.
\qed\end{example}

\subsection{Explore Sets of ASMs}\label{sec:ex}

\begin{table}[t]
\[
\renewcommand\arraystretch{1.4}
\begin{array}{| c || c | c |}\hline
& \multicolumn{1}{c|}{\mbox{Explore set $\Gamma(X)$}}
  & \mbox{Exploration order $\prec_X$}\\\hline\hline
0 & j\neq n,\; j=n \wedge i+1\neq n& \\\hline
1  & j\neq n,\; j=n \wedge i+1\neq n,\;i+2 & j=n \wedge i+1\neq n  ~\prec_X~i+2\ \\\hline
2 & \multirow{2}{*}{$\begin{array}{c} j\neq n,\; j=n \wedge i+1\neq n,\;\\F(i)>F(j),\;j+1\end{array}$}
& \multirow{2}{*}{$\begin{array}{c} j\neq n~\prec_X~\\F(i)>F(j),\;j+1 \end{array}$} \\\cline{0-0}
3 & & \\\hline
\end{array}
\]
\caption{Explore sets (omitting subterms) and an exploration order for the cases of the sorting example
 (Figure~\ref{fig})
shown in Table~\ref{tab:T}.}\label{tab:G}\label{tab:L}
\vspace*{3mm}\hrule
\end{table}

As explained earlier, one can easily model partial functions by using a special ``undefined'' value.
The problem is that we need to model the case when an algorithm calls
such a function, but the function never informs the algorithm that it is undefined for the arguments in question.
This situation should entail that the algorithm ``stalls''.
See Section~\ref{sec:part}.
But then it is crucial that only intended locations are explored during an emulation,
something that was irrelevant to the original study.
For this reason, we make explicit now which locations are actually explored by an ASM.
For this reason, too, it
behooves us to refine the ASM Theorem so that every classical algorithm
can be emulated by an ASM that does not explore locations with undefined values,
unless the algorithm also does.
This refinement will be undertaken in Sections~\ref{sec:exp} and~\ref{sec:le}.

Let $\Gamma(X)\subseteq T$ denote that set of critical terms that are actually explored by the algorithm
at state $X$, so as to determine
how to continue the computation.
We will call $\Gamma(X)$ the \emph{explore set} of $X$.
For ASMs, $\Gamma$ would include the actual tests performed, and the terms needed for the actual assignments.
It may be defined inductively in the following fashion for ASMs:
\begin{equation}\label{eq:gamma}\setlength{\arraycolsep}{3pt}
\begin{array}{lcl}
\Gamma_{[P_1 \;\|\; \cdots \;\|\; P_n]}(X) &=& \bigcup\nolimits_j \Gamma_{P_j}(X)\\
\Gamma_{\mbox{\bf \small if} \; C\;\mbox{\bf \small then}\,P}(X) &=&
 \{C\}  \cup
\begin{cases}
\Gamma_P(X)~~ & \mbox{if $X\models C $}\\
\emptyset & \mbox{otherwise}
\end{cases}\\
\Gamma_{f(s_1,\dots,s_n) := t}(X) &=&  \{s_1,\ldots,s_n,t\}\;,
\end{array}
\end{equation}
where $X\models C$ means that Boolean condition $C$ holds true in $X$.
Thus, $\Gamma(X)$ includes all the conditions that are actually tested when in state $X$,
and all the terms appearing in updates that are actually performed.%
\footnote{If one chooses to leave Boolean operations out of the vocabulary
of states, as suggested in footnote~\ref{Bool}, then $\Gamma$ should contain the atomic predicates appearing in
the conditions,
without connectives.}
In addition, $\Gamma$ always includes \True{} and \False, and all subterms of its members,
since the latter need to be evaluated before the locations denoted by the above terms can be accessed.
See Section~\ref{sec:vary} for a discussion of the different aspects of exploration.
%see Section~\ref{sec:clash} for ASMs in which assignments may clash.

\begin{example}
Table~\ref{tab:G} gives the explore sets $\Gamma(X)$ for our sorting program.
For example, if $i+1=j=n$ in output state $X$ (row 0), then $\Gamma(X) $ is  just
$\{j\neq n,\, (j=n)\wedge (i+1\neq n)\}$, plus their subterms.
\qed\end{example}

In what follows, we fine-tune the Bounded Exploration Postulate, by making explicit the possibility that
only a subset of the critical terms may be needed in any particular situation.

\section{Exact Exploration}\label{sec:exp}\label{sec:be}

Before delving into an analysis of the exploration of states, we should
visualize for ourselves how an algorithm goes about retrieving data from its current state
and storing updated information for what will be the next state.

\subsection{Executor Model}\label{sec:model}

One should distinguish between an algorithm proper and its state.
An algorithm is a finite collection of instructions of some sort,
``run'' by an ``executor''.%
\footnote{The latter is the ``computing agent'' $L$ of \cite{Rogers}.}
An example of an algorithm is the method of long division,
in which case the executor might be a pupil who has fully mastered the rules.

In general, a state may be a finite or infinite object.
It must be kept in some form,
like scratch paper used in the process of dividing, and
the multiplication table that needs to be looked up.
Furthermore, classical algorithms are deterministic, so there is no need for the algorithm to guess what
to do next.

Static information, including native operations and methods, is a fixed part of an algorithm's state.
When, for example, a long division algorithm needs to look up a small multiplication table,
it turns to the state.  When it needs to add two numbers, it also appeals to the built-in addition operation
provided by the state.
In addition, all dynamic information is stored in the state.

The executor of the algorithm, on the other hand,
need remember nothing about previous states, since all that is relevant is available in the current state.
The executor acts on states according to the instructions contained in the algorithm being executed.
By definition, the state contains all information that, in addition to the program, determines the
future behavior of the algorithm.
The executor can take a ``lunch break'' between steps and continue the process exactly where it was left off.
History does not matter at all.
The state of a Turing machine, for instance, must include the information in
its instantaneous description (that is, the tape contents, head position, and
internal state), plus operations for reading, writing, changing internal state, and moving the head.
That is what is required to determine the next state.

At the beginning of a step, all states look the same to the executor.
That is why exploration of a state must always start the same way.
It is possible that the executor applies a fixed update set without ever querying the state.
But this is a degenerate case.
In general, the executor requires some information from the state to complete the next step.
The executor consults the state by posing a batch of questions
for the ``state manager'' to answer.
Questions for which the executor knows the answer need not be asked,
but we are not precluding that possibility.

What kind of questions?
Note that it is of no help if the state manager displays an actual element to the algorithm executor.
This follows from the Abstract State Postulate.
The executor understands only objective things, things that do not depend on the particular implementation.
In the original framework, the only objective thing was equality:
equality meant the same to the state as to the executor.
If we want to eliminate the dependence on full equality, we need to change that aspect of ASMs.
We still presume that the executor and manager have
an agreed-upon interface.
%\footnote{An ``interface'' is some convention for communication.
%As in the ordinary, interactive case of \cite{OrdI}, all communication may be reduced to a series
%of queries and responses.
%We do not impose a ``one question at a time'' discipline.}
In particular, we assume that they both understand immediately and give the same import to the truth value constants $\True$ and $\False$.
More generally, there could be some other small set of agreed-upon values upon which
communication between algorithm and state can be based
and for which the vocabulary includes distinct self-denoting (static nullary) symbols.
See Section~\ref{sec:eq}.

\subsection{Refined Analysis}\label{sec:refine}

As already explained, deciding which locations to explore is part of the behavior we are now interested in.
If an algorithm acts differently on different states,
either in the sense of exploring different terms or
in the sense of performing different updates,
then it clearly must \emph{first} find something that distinguishes them.
So
we certainly want both (\ref{eq}) and (\ref{eq2}) above.
Furthermore,
if the behaviors of the algorithm in states $X$ and $Y$ differ, then
that must be made evident from the part of the state that is explored both in $X$ and in $Y$.
Accordingly, what we really want is more like the following:
\begin{eqnarray}
\label{eq:cap}
&\mbox{\it if } X =_{\Gamma(X)\cap\Gamma(Y)} Y \mbox{\it\  then } \Gamma(X)=\Gamma(Y)\\[-0pt]
&\mbox{\it if } X =_{\Gamma(X)\cap\Gamma(Y)} Y \mbox{\it\  then } \Delta(X) = \Delta(Y)\;.
\end{eqnarray}

A bit of notation.
For a set $V$ of algorithm states,
let $\Gamma(V)$ be short for the \emph{shared} explore terms $\bigcap_{X\in V}\Gamma(X)$.
We will say that $V$ is \emph{agreeable} if all states in $V$ agree on
the values of all their shared explore terms, that is, if
$X =_{\Gamma( V)} Y$ for all $X,Y\in V$.
It stands to reason that sets of agreeable states engender uniform behavior,
because the algorithm has no way of distinguishing between them.

We defer until Section~\ref{sec:op} consideration of the possibility that operations may be undefined for some arguments, in which case some terms may not have any value at all in a given state.

For an algorithm to proceed, it needs to communicate with its state $X$,
as described above. In particular, the algorithm may need to learn information that distinguishes $X$ from other states for which it behaves differently.
To this end, the algorithm evaluates---in some order---a finite collection of terms over $X$ and learns their values.
In addition, in order to produce updates, the algorithm evaluates a finite collection of additional terms, the values of which need not be actually observed.

Think of it this way:
An algorithm starts out agnostic about the nature of the current state.
It may begin by performing some updates, but only such updates as are not contingent on
state.
If (but not only if) any aspect of its behavior is contingent, then
the next thing it does is evaluate some set $G$ of critical terms.
So, in all events, $\Gamma(\St)$ includes $G$.
If all states in $\St$ happen to agree on $G$, then, at this point, either no further exploration is undertaken,
and all states in $\St$ must have the same behavior,
or else in every state of $\St$ the algorithm goes on and evaluates some additional set
$G'$ of terms, so as to distinguish different behaviors.
If not all states agree on $G'$, then, depending on the truth values of terms in $G'$,
the algorithm proceeds differently in the different cases.
There may also be some shared behavioral aspects that may be performed regardless of the outcome of evaluating $G'$.
If the different behaviors are still not fully distinguished by $G'$, then an additional
set $G''$ of terms is called for, for each set of answers to $G'$.  And so on.
Note that for agreeable $\St$, $\Gamma(\St)$ would also include $G'$, and all states in $\St$ would also agree on $G'$,
and---in the final analysis---the behavior of the algorithm would be uniform for all states in $\St$.

The precise order of exploration need not be fixed for a given state, but some partial order
is dictated by the possible behaviors.

In general, in a given state $X$, if a conditional ASM statement $\If C \Then P$ is executed
and the test $C$ is true,
then the terms in $C$ are explored before, or together with, those in $P$.
But we cannot simply derive the exploration order from the conditionals in the program,
making conditions in $C$
smaller than any new terms in $P$.  For example, we might have an assignment
$\If d \Then \If b \Then x:=d$, in which case $d$ needs to be explored before $b$,
but when placed in parallel with
$\If b \Then \If d \Then x:=c$, $b$ and $d$ can be explored at the same time.
Instead, we put all terms of the top-level conditions and assignments of components of
a parallel statement at the bottom of the ordering, followed by contributions from the relevant cases of the conditionals.

\begin{example}\label{ex:6}
Consider the following ASM program, in the expanded form described at the end of Section~\ref{sec:prog}:
\[
\begin{array}{lll}
[ & \If d \Then \If c \Then \If b \Then s := x &\|\\
&\If d \Then \If \neg c \Then t := x &\|\\
&\If d \Then \If \neg b \Then s := y &]\;.
\end{array}
\]
Clearly, $d$ must be explored first off, since nothing more transpires when $d$ is false,
while further tests are necessary when $d$ is true.
Suppose the latter is the case.
Then $b$ and $c$ must both be explored, though the order in which that occurs does not matter.
Of course, $x$ and/or $y$ are only explored after it becomes clear that
the relevant case holds.
(Note that the algorithm need not ascertain the values of $x$ and $y$;
those locations are used only for the purpose of transferring their contents to
locations $s$ and $t$.
See Section~\ref{sec:access}.)

We have the following (omitting some self-evident subscripts):
\begin{itemize}
\item $\Gamma(X)=\{d,c,b,x\}$ and $\Delta(X)=\{s\mapsto \val{x}{}\}$ whenever $X\models d,c,b$;
\item $\Gamma(Y_0)=\{d,c,b,x,y\}$ and $\Delta(Y_0)= \{t\mapsto\val{x}{},s\mapsto\val{y}{} \}$ whenever $Y_0\models d,\neg c,\neg b$;
\item $\Gamma(Y_1)=\{d,c,b,x\}$ and $\Delta(Y_1)=\{t\mapsto\val{x}{}\}$ whenever $Y_1\models d,\neg c,b$;
\item $\Gamma(Y_2)=\{d,c,b,y\}$ and $\Delta(Y_2)=\{s\mapsto\val{y}{}\}$ whenever $Y_2\models d,c,\neg b$;
and
\item $\Gamma(Z)=\{d\}$ and $\Delta(Z)=\emptyset$ whenever $Z\models \neg d$.
\end{itemize}
In a state $X$ with $d$, $c$, and $b$ true,
we must have $d$ explored before $b$ or $c$, which are both explored before $x$ is, while $y$ is not examined.
But whether $c$ is explored before $b$, after $b$, or simultaneously with $b$ is immaterial.
This is because once $d$ is true, $b$ must be examined regardless of the truth of $c$,
so as to determine if $y$ needs to be updated by the third conditional of the program.
\qed\end{example}

This \emph{order} of exploration will be captured in what follows by a ``causality'' order $\prec_X$ on the
explore terms $\Gamma(X)$ of states $X$.

\subsection{Refined Postulate}

With these considerations taken into account,
the refined exploration postulate, replacing Postulate~\ref{BE}, is as follows:

\renewcommand{\thepostulate}{IIIe}
\begin{postulate}[Exact Exploration]\label{EE}
An algorithm with states $\St$ over vocabulary $\U$
determines, for each state $X\in\St$,
a finite explore set $\Gamma(X)$ of ground terms over $\U$
such that the following three properties hold:
\begin{itemize}
\item {\bf Determination.}
For all states $X,Y\in\St$,
if $X =_{\Gamma(X)} Y$, then $\Delta(X) = \Delta(Y)$.
\item {\bf Discrimination.}
For each state $X\in\St$ there is a partial order $\prec_X$ of $\Gamma(X)$ such that for
every state $Y\in\St$ and for any $t\in \Gamma(X)\setminus\Gamma(Y)$,
% if any,
there is a Boolean term $s \in \Gamma(X)$ that takes on opposite truth values in $X$ and $Y$ and such that $s \prec_X t$.
\item {\bf Limitation.}
The set $\bigcup_{X \in\St} \Gamma(X)$ of all explore terms is finite.
\end{itemize}
\end{postulate}
\noindent
By ``opposite'', we mean that in one state $s$ has the same value as \True, and in the other, as \False.

Determination says that if two states $X$ and $Y$ agree on the values of
the explored terms $\Gamma(X)$,
then the changes that need to be made from each to get to their next states are the same.
This part is analogous to the original version of the third postulate,
except that the set of critical terms is localized so as to depend on the state $X$.
The second part (Discrimination) ensures that an algorithm also determines which locations are to be explored before actually exploring them, so if a term $t$ is explored in $X$ but not in $Y$,
then that distinction depends on some previously explored term $s$.
The fact that algorithmic behavior is finitely describable is captured by the last part (Limitation), which
states that only finitely many terms need to be mentioned to fully characterize what locations are to be explored
and what changes are to be made.
The original postulate likewise insisted that the set of all critical terms is finite.

\begin{example}
The rightmost column of Table~\ref{tab:L} shows the partial order in which the explore terms are examined by the sorting algorithm of Figure~\ref{fig}.
\qed\end{example}

\begin{remark}\label{rem:3}
Equation~(\ref{eq:cap}) does not suffice for Discrimination.
Consider a pseudo-algorithm
that ``magically'' chooses to execute
\[
[\;\If b \Then [b := b ~\|~ c := c]\quad
 \;\|\; \quad \If \neg b \Then d := d\;] \;,
\]
whenever $d$ is true,
and otherwise executes
\[
[\;\If c \Then [b := b ~\|~ c := c]\quad
 \;\|\; \quad \If \neg c \Then d := d\;] \;,
\]
and has the following three initial states:
$X\models \neg b,\neg c,d$; $Y\models b,c,\neg d$; and $Z\models b,\neg c,\neg d$.
We have $\Gamma(X)=\{b,d\}$, $\Gamma(Y)=\{b,c\}$, and $\Gamma(Z)=\{c,d\}$.
It is easy to verify that this algorithm satisfies the simple version (\ref{eq:cap}).
But it does not meet the requirement of having
strictly smaller discriminating terms.
To see this, note that one must have $b \prec_X d$ to discriminate between $X$ and $Y$,
but must also have $d \prec_X b$, because $d$ is the only
Boolean that discriminates between $X$ and $Z$.
In fact, without the ``magic'', an algorithm would have no way of
knowing whether to start by exploring $b$ or $c$.
\end{remark}

\begin{definition}[Exacting Algorithm]
An algorithm
satisfying Postulates \ref{P1}, \ref{P2}, and \ref{EE}
will be called \emph{exacting}.
\end{definition}

The following property is certainly to be expected if exploration is an aspect of behavior of an algorithm:

\begin{lemma}
For any exacting algorithm and isomorphic states $X$ and $Y$,
we have $\Gamma(X)=\Gamma(Y)$.
\end{lemma}

\begin{pf}
Isomorphic states agree regarding the truth values of Boolean terms.
So  there can be no $s$ to discriminate between  $\Gamma(X)$ and $\Gamma(Y)$,
as required by Discrimination.  Hence,  $\Gamma(X)=\Gamma(Y)$.
\end{pf}

\subsection{Exact Exploration is Bounded}\label{sec:e}

Since, as in the original study, we are aiming at a universal formalization of algorithms,
we should expect the same processes to obey our new Exact Exploration Postulate
as fulfill the original Bounded Exploration Postulate.
The difference stems from the
fact that the critical terms that are actually examined may depend on which state the algorithm is currently examining, an aspect of behavior captured by the new postulate, but not by the original.

\begin{theorem}\label{thm:e}
Every exacting algorithm is also classical,
and every classical algorithm can be equipped with explore sets so as to be exacting.
\end{theorem}

\begin{pf}
To see that Exact Exploration (Postulate~\ref{EE}) implies Bounded Exploration (Postulate~\ref{BE}),
let $T=\bigcup_{X\in\St} \Gamma(X)$ be all the explore terms of an algorithm satisfying the former.
By Limitation, $T$ is finite;
by Determination, $T$ determines behavior as required by Postulate~\ref{BE}.

For the other direction,
we let $\Gamma(X)=T$, for all states $X\in\St$, where $T$ is the algorithm's finite set of
critical terms per Postulate~\ref{BE}.
It is straightforward to see that all parts of Postulate~\ref{EE}
are fulfilled:
Determination and Limitation by Bounded Exploration,
and Discrimination, vacuously, with any order $\prec_X$.
\end{pf}

Since an exacting algorithm is classical,
it has an emulating ASM by Theorem~\ref{thm:asm}.
This, however, is insufficient for our stated purposes,
since that ASM might explore any location given by $T$, not just
those of the current state given by $\Gamma(X)$.
What we show in the next section is that there is an emulating ASM
that always restricts exploration to $\Gamma(X)$,
so the ASM satisfies a more refined notion of equivalence
(Definition~\ref{def:eq} below) than
used in Theorem~\ref{thm:asm}.

We also just showed that every classical algorithm can be made exacting, but in a rather uninteresting way: all states share the same explore set.
Explore sets, however, can be much more informative.
Accordingly, what we consider next are exacting algorithms with non-uniform $\Gamma(X)$.

\section{Exacting Algorithms}\label{sec:le}

Our goal now is to show that every exacting algorithm is
equivalent to some abstract state machine, where the notion of equivalence pays attention to both
transitions and exploration.
This will enable us to refine the Abstract State Machine Theorem to take
into account which locations in a state are actually explored.

\subsection{Refined Equivalence}\label{sec:eqv}

There are many senses in which one may say that two algorithms are equivalent.%
\footnote{See \cite{PITS} for a discussion of the slippery notion of equivalence of algorithms.}
We are interested here in behavioral equivalence of two transition systems.
In addition to updating states 
in the same fashion,
we want behavioral equivalence of algorithms to also mean that the same critical terms are explored
in each state. Accordingly, for the purposes of this paper, we may define ``algorithm equivalence''
for exacting algorithms as follows:

\begin{definition}[Equivalence]\label{def:eq}
Two exacting algorithms $\cal P$ and $\cal Q$
are \emph{equivalent} if they operate over the same states $\St$, have the same initial states $\In$ and terminal states $\Ot$,
and
provide exactly the same explore sets and update sets, that is, if,
for all states $X\in\St$,
$\Delta_{\cal P}(X) = \Delta_{\cal Q}(X)$
and $\Gamma_{\cal P}(X) = \Gamma_{\cal Q}(X)$.
\end{definition}

Except to the extent that the order in which locations are explored might affect what is actually explored in a given state,
we do not care about the precise order of exploration,
nor about the number of times a location is accessed.
Should one be interested in those additional aspects of an algorithm's behavior,
one ought to lower the level of abstraction and decompose the
individual steps to make those features explicit.%
\footnote{Were there a need to consider exploration orders as part of the behavior of algorithms, and therefore to require equivalent algorithms to agree as to the orderings, Postulate~\ref{EE} would need to be strengthened in two ways. First, if two states produce different orderings, then any such difference should be ``caused'' by a difference of values at some location that was explored earlier.  Second, the orderings should be constrained to look like the intuitive picture in Section~\ref{sec:refine}: An initial block $G$ of terms, followed by a second block $G'$ (that can depend on the answers to the first block), etc.
% That is, the partial order amounts to a linearly ordered sequence of blocks.  Formalizing these ideas, one could develop an analogous theory, and in particular
Then one can prove that algorithms of this sort can be emulated by ASMs---including matching the orderings.
To make the frequency of exploration of locations also part of behavior would require the use of multisets for $\Gamma$.}

\subsection{Uniformity}

We will say that a set $V$ of states is \emph{uniform} if all states in $V$ have the same
explore set, that is, if
$\Gamma(X) = \Gamma(Y)$ for all $X,Y\in V$.
Recall from Section~\ref{sec:refine} that $\Gamma(V)=\bigcap_{X\in V}\Gamma(X)$
is the set of their shared explore terms, and that
$V$ is said to be agreeable when all states in $V$ agree on the values of $\Gamma(V)$,
that is, if $\val s X = \val s Y$ for all states $X,Y\in V$ and terms $s\in\Gamma(V)$.

When $V$ is agreeable, then it should also be uniform:

\begin{theorem}\label{thm:n}
For any exacting algorithm, agreeability of a set of states implies its uniformity.
\end{theorem}

\begin{pf}
By contradiction, suppose that, despite $V$'s agreeability,
not all states in $V$ have the same explore set.
Without loss of generality, let $t\in\Gamma(X)$ be a minimal explore term for some $X\in V$ that is not also
an explore term for all other states in $V$ (minimal with respect to $\prec_X$),
and let $Y\in V$ be a state such that $t\notin\Gamma(Y)$.
By Discrimination, there is an $s\in\Gamma(X)$ such that $s \prec_X t$ and with
different truth values in $X$ and $Y$.
By agreeability, $s\notin\Gamma(V)$.
But then $s$ must be a smaller choice of an explore term for $X$ than is $t$,
since perforce $s\notin\Gamma(Z)$ for some $Z\in V$.
\end{pf}

%\begin{remark}
Consider, again, the example in Remark~\ref{rem:3}.
Since the intersection of the three different explore sets is empty,
it cannot be that agreeability, which holds vacuously for those states,
always implies uniformity.
It must be, then, that the explore sets of the ``magic'' algorithm in question are not discriminating.
%\end{remark}

By Determination, we also have the following:

\begin{corollary}\label{cor}
For any exacting algorithm, agreeability of a set $V$ of states
implies that
$\Delta(X) = \Delta(Y)$ for all $X,Y\in V$.
\end{corollary}

\begin{theorem}\label{thm:nachum}
For any classical algorithm equipped with explore sets satisfying the Determination and Limitation clauses of Postulate~\ref{EE}, if every agreeable set of states is uniform,
then the algorithm is exacting.
\end{theorem}

\begin{pf}
For each $X\in\St$, we define a partial order $\prec_X$ on $\Gamma(X)$.
Explore terms that are shared by all states are smallest, because they
are always needed.
Next come
those terms that are shared by all states that agree with $X$ on the values of the lowest tier,
$\Gamma(\St)$, of terms.  And so on.
Thus, the ordering $\prec_X$, as a set of ordered pairs, is
$L_X(\St)$, where $L_X(V)$ is  an ordering that discriminates $X$ from other
states in $V$.  When  $V$ is uniform,
$L_X(V)  :=  \emptyset$;
otherwise,
\[
L_X(V)  :=
(\Gamma(V)\times(\Gamma(X)\,\setminus\,\Gamma(V)))\, \cup\,
 L_X(\{Y\in V\st Y=_{\Gamma(V)} X\}) .
\]

This recursion is bound  to terminate, because $\Gamma(X)\setminus\Gamma(V)$ gets continually smaller.
To see why, note that $X\in V$ always,
so $\Gamma(V)\subseteq\Gamma(X)$.
When $V$ is not uniform, it cannot be agreeable,
so there is an $s\in\Gamma(V)$ over which states in $V$ disagree.
But, by construction, all of $V$ agrees on all terms in the previous $\Gamma(V)$.

To show Discrimination, consider any $t\in\Gamma(X)\setminus\Gamma(Y)$ for a $Y\in\St$.
Initially, $t\notin\Gamma(V)=\Gamma(\St)$, whereas $t\in\Gamma(V)=\Gamma(X)$
at the end of the recursion, so $Y$ is not in the final argument $V$.
At the point when $Y$ is removed from $V$, there must be an $s\in\Gamma(V)$
that discriminates between $X$ and $Y$.
By construction, $s \prec_X t$.
\end{pf}

It follows that the Discrimination requirement is equivalent to
``agreeability implies uniformity''.

\begin{example}
For the program in Example~\ref{ex:6}, and for $X\models d,c,b$, the construction in the above proof yields
$d \prec_X b, c \prec_X x$.
\qed\end{example}

\subsection{ASMs are Exacting}

By the Theorem~\ref{thm:e} of Section~\ref{sec:e},
any exacting algorithm also satisfies Bounded Exploration,
and any classical algorithm satisfies Exact Exploration, when the
explore terms of every state are taken to be all critical terms. %, unordered.
A classical algorithm can always be simulated by many ASMs,
which may differ from one another in terms of what tests need
to be performed in any given state.
We show now that the precise explore terms of ASMs
satisfy our refined postulate.

\begin{theorem}
Every (clash-free) ASM program is an exacting algorithm.
\end{theorem}
%See Section~\ref{sec:clash} for the case where the assignments in the ASM may clash.
\noindent
See Section~\ref{sec:clash} for when assignments may clash.

\begin{pf}
We know from the original study that Postulates~\ref{P1}--\ref{P2} hold for ASMs.
So, we only need to show that explore terms, as defined for ASMs
in Section~\ref{sec:ex}, satisfy Postulate~\ref{EE}.

Limitation clearly holds, since ASM programs are finite and all terms in any $\Gamma(X)$ appear in the program.

Furthermore,
if $X=_{\Gamma(X)}Y$, then, in particular, all tests performed by the algorithm have the same outcome in $Y$ as in $X$, since they are included in $\Gamma(X)$.
So $\Gamma^+(X)=\Gamma^+(Y)$, and, hence, $\Gamma(X)=\Gamma(Y)$, as demanded by Determination.

Appealing to Theorem~\ref{thm:nachum}, we show Discrimination for ASMs by showing that
agreeability of a set of states implies its uniformity.
This follows by induction on the syntax of ASM programs:
Assignments contribute uniform explore sets always.
Parallel composition preserves uniformity, assuming agreeability.
Conditions of \textbf{if} statements are always in the explore sets,
so, by agreeability, either all states get the contributions of the then-branch,
or none do.
\end{pf}
%Note that the algorithm (\ref{eq:gamma}) for determining the explore set of an ASM works
%equally well for a set of states $V$ as for a single state $X$.
%It returns $\Gamma(V)$, understanding $V\models C$
%as meaning that every state $X$ in $V$ satisfies condition $C$.
%If $V$ is agreeable, then whenever some state in $V$ does not satisfy $C$, none do.
%So the computation of $\Gamma(V)$, for agreeable $V$, proceeds precisely as
%does the computation of $\Gamma(X)$ for any $X\in V$.
%It follows that $V$ is also uniform.

%\subsection{Exacting Algorithms have Programs}
\subsection{Refined ASM Theorem}

The differences between this paper and the original study
have no impact on the following observation \cite[Lemma~6.2]{Gurevich00},
other than to localize critical terms:

\begin{lemma}\label{lem} %[{\cite[Lemma~6.2]{Gurevich00}}]
For every exacting algorithm,
if $f(\bar a)\mapsto b$ is an update in $\Delta(X)$ for some state $X$,
then there are terms $t$ and $\bar s$ in $\Gamma(X)$ such that
$\val{t}X = b$ and
$\val{s_i}X = a_i$ for each $a_i$ of $\bar a$.
\end{lemma}

It follows that for any update $f(\bar a)\mapsto b$ in $\Delta(X)$
there is an assignment statement $f(\bar s):=t$ that has the
desired effect and which is constructed only from explore terms in $\Gamma(X)$ (and the
symbol $f$).

\begin{figure}
\begin{center}
\includegraphics[scale=1.2]{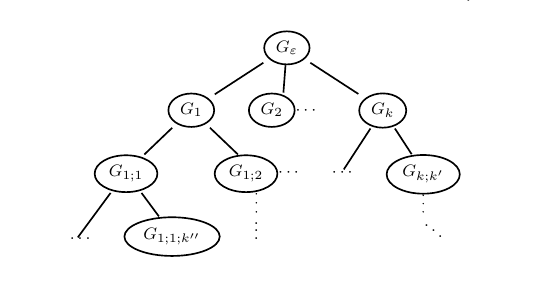}
\caption{Exploration tree.}\label{tree}
\end{center}
\vspace*{-3pt}\hrule
\end{figure}

The result we have been seeking is the following:
\begin{theorem}[Refined ASM Theorem]\label{thm}
Each exacting  algorithm has an equivalent (clash-free) ASM.
\end{theorem}
\noindent
Section~\ref{sec:clash} addresses the case when assignments may clash.

\begin{pf}
Let $\St$ be the set of states of algorithm $\cal Q$, 
with
explore set $\Gamma$ and update set $\Delta$.
The equivalent ASM $\cal P$ operates over the same states $\St$.
Its program is $P(\St)$, defined recursively as follows:
\[
P(V) ~=~
\left\{\!\!\!\begin{array}{ll}
[\;] & \mbox{if $V=\emptyset$}\\[1ex]
[\If C_1\Then P(V\restrict C_1)\|\cdots\| \\[3pt]
\If C_k\Then P(V\restrict C_k)]
& \mbox{if $V$ is not agreeable}\\[1ex]
 \If C \Then R & \mbox{if $V$ is agreeable}
\end{array}\right.
\]
where
$C_1,\dots,C_k$ are mutually-exclusive
Boolean conditions for partitioning the states in $V$
according to all possible truth assignments to Boolean terms in $\Gamma(V)$
(those terms whose value is not agreed upon by all states in $V$),
and $V\restrict C_i$ are those states $X\in V$ for which $X\models C_i$.
When $V$ is agreeable,  there is no partitioning, as
all states in $V$ have the same explore and update sets
(Theorem~\ref{thm:n} and Corollary~\ref{cor}).

In the agreeable case, the program is of the form $ \If C \Then R$.
The test $C$ is a conjunction of all the (unexplored) Boolean terms $c\in\Gamma(V)$, or their negations $\neg c$,
depending on whether $V\models c$ or $V\models\neg c$.
The purpose of $C$ is to ensure that all explore terms are indeed explored.
If there are none, then $C$ is vacuously true and the condition may be omitted entirely.
The program $R$ is a parallel collection of assignments for all the updates in
$\Delta(X)$, for any one state $X\in V$,
the existence of which follows from Lemma~\ref{lem} and the uniqueness of which follows from Corollary~\ref{cor}.

To see why the recursion terminates, consider
Figure~\ref{tree}, depicting the above construction.
The root $G_\varepsilon$ contains $\Gamma(\St)$, those
terms that are invariably explored.
Every other node corresponds to a call $P(V\restrict C_i)$,
and contains the next level of discriminating terms,
namely, $\Gamma(V\restrict C_i)\setminus\Gamma(V)$.
The second layer of the tree is populated by finitely many nodes, each for a
possible combination of values for the terms in $G_\varepsilon$, with $G_1=\Gamma(\St\restrict C_1)\setminus\Gamma(\St)$,
$G_2=\Gamma(\St\restrict C_2)\setminus\Gamma(\St)$, and so on.
Since every node contains a nonempty set of explore terms from the finite set
$\bigcup_{X\in\St}\Gamma(X)$, none of which appear above it in the tree,
all paths are finite.

It can be seen that this program has the same explore and update sets as does the original algorithm, since
for any $X\in\St$ the appropriate statement $ \If C \Then R$ is executed.
\end{pf}

\begin{example}
Returning to Example~\ref{ex:6},
we have $\Gamma(\St)=\{d\}$,
and $\Gamma(\{X,Y_0,Y_1,Y_2\})=\{d,c,b\}$.
The above construction yields
\[
\begin{array}{llllrr}
[ & \If d \Then &
[
& \If c \wedge b \Then s := x & \\
&&\|& \If c \wedge \neg b \Then t := x & \\
&&\|& \If \neg c \wedge b \Then  s := y & \\
&&\|& \If \neg c \wedge \neg b \Then  [t:=x \| s:=y] ~~ ]\\
\| & \If  \neg d \Then & [~] & &]
\end{array}
\]
which is equivalent to the original.
\qed\end{example}

For reasons similar to those for Theorem~\ref{thm:n}:

\begin{lemma}\label{lem:less}
For every exacting algorithm, set of states $V$, and state $X\in V$, it is the case that
for each term $t\in\Gamma(X)\setminus\Gamma(V)$, there is
some term $s\in\Gamma(V)$ such that $s \prec_X t$.
\end{lemma}

\begin{pf}
It suffices to show this for all \emph{minimal} elements $t\in\Gamma(X)\setminus\Gamma(V)$.
Since $t\notin\Gamma(V)$, there must be a $Y\in V$ such that $t\notin\Gamma(Y)$.
Discrimination implies
there is some $s\in\Gamma(X)$ such that $s \prec_X t$.
Since $t$ is minimal, it must be that $s\in\Gamma(V)$.
\end{pf}

%\subsection{Some Observations}
%
%Some points to keep in mind follow:
%
%\begin{lemma}\label{lem:eq}
%Given Eq.~(\ref{eq:cap}),
%the following are equivalent, for all states $X,Y\in\St$:
%\begin{enumerate}
%\item\label{a} $X =_{\Gamma(X)\cup\Gamma(Y)} Y$;
%\item\label{b} $X =_{\Gamma(X)} Y$;
%\item\label{c} $X =_{\Gamma(X)\cap\Gamma(Y)} Y$.
%\end{enumerate}
%\end{lemma}
%
%\begin{pf}
%%The only non-trivial implication is
%%\yg{The first part of the proof is confusing. We may want to replace it with something innocuous like ``We have''.}
%By inclusion,  (a) implies (b) implies (c).
%Equation~(\ref{eq:cap}) demands that
%$
%X =_{\Gamma(X)\cap\Gamma(Y)} Y
%\Rightarrow
%\Gamma(X)=\Gamma(Y)$,
%from which it follows that
%$\Gamma(X)\cup\Gamma(Y) = \Gamma(X) = \Gamma(X)\cap\Gamma(Y)$.
%So (c) implies (a).
%\end{pf}

\section{Partiality}\label{sec:part}

The refined ASM Theorem of  the previous section
allows us to model truly partial functions and relations.

\subsection{Partial Equality}\label{sec:eq}

Equality was sacrosanct in the original study, in
that each and every state of an algorithm is endowed with the logical equality relation between arbitrary elements of its base set.
This may be unrealistic, however.
An algorithm might need to divide arbitrary real numbers, but not have the ability to test for zero.
So, suppose, instead, that equality is internal to states. In other words, the vocabulary itself includes a
symbol ``='' for equality.
Normally, equality evaluates to one of two Boolean values, \True{} and \False.
But, in some cases, an equality test might realistically ``hang'' and not return any answer.
In other cases, an equality test might return \Undef, thereby explicitly informing
the algorithm that no definite answer is forthcoming.
The difference is that hanging is more insidious---the computation gets stuck in a catatonic limbo,
while an explicit undefined answer is more like an error message---for the algorithm
to handle as it sees fit.
In any case, we do insist that the \emph{defined} (non-hanging, not-\Undef) values of ``='' agree with true equality.
With this flexibility, for states to be isomorphic, the isomorphism
must, of course, also respect the provided equality relations.

Consider the following scenario:
The base set of states includes expressions (like definite integrals) that represent real numbers.
A natural equality relation on such expressions (in contradistinction to identity of expressions)
deems two expressions ``equal'' iff they represent the same number.
States might implement a partial version ($\approx$) of this equivalence,
which need not be transitive.
Furthermore, it could be that a test $s\approx t$ yields \False,
whereas $t\approx u$ yields \True,
yet when asked about $s\approx u$, no answer is forthcoming,
though the truth of the matter must be that $s\not\approx u$.
%{Y: Concerning partial equality, the situation is slightly more complicated. There are three levels:
%1. Genuine reals. 2. The various expressions (integrals, etc.) that the expert use. 3. The expression in our language that evaluate to entities of type 1 or 2. The equality may be partial only on entities of type 2.}

\subsection{Partial Operations}\label{sec:op}

All function symbols were total in the original study.
With the machinery developed here, one need not insist that states interpret each function symbol fully.

Instead, let a state be a structure that allows for truly partial functions.
Whenever an algorithm explores an undefined location,
the computation hangs.
One may, of course, have ``error'' values, such as \Undef,
as used extensively in the ASM literature,
but these are in fact ``defined values'', which do not cause irrecoverable execution collapse.
In cases where the domain of definition is undecidable, it may not
be accurate to fill those non-domain points with \Undef,
especially if one wants an implementation whose initial states are computable.

We need to give semantics to terms involving partial functions.
The sensible choice is to make a term ``truly'' indeterminate
if any of the locations indicated by any of its subterms is indeterminate.
We consider that accessing a location
$f(\bar x)$ hangs when $\bar x$ is not in the domain of definition of $f$,
extended to include all known $\Undef$ cases.
All operations, including equality and Boolean operations,
are \emph{strict} in the sense that if any operand is indeterminate, then that operation
is also.
By convention, we will indicate such circumstances
by writing $\val{t}{X}=\hole$, with {\hole} standing for ``no value at all'' (not even \Undef),
whenever term $t$ does not evaluate in state $X$ to an element in its base set.
So,
% better displayed
\[\val{f(\dots,\hole,\dots)}{X} = \hole \]
for all states $X$ and operations $f$, regardless of argument values for the terms
of the ellipses.
This is in contrast with the explicit undefined value for which the test
$\Undef=\Undef$ returns \True.

In general,
should an algorithm ask a state $X$ for the value
of an operation at a point where it is undefined,
the state will be unable to answer, and the program must hang.
Thus, any attempt to access an undefined value
means that there will be no next state, and---what is worse---no
way for an observer to know that the algorithm is stuck, that there is no point waiting for an answer.
We write $\Delta(X)=\hole$.

For example, if the base set includes programs in some language and states include a
black-box operation that interprets programs on inputs,
any test whether the result of applying a program to given inputs yields a particular output
should hang whenever the program does not halt for those inputs.
%\footnote{This situation is analogous to the simple form of interaction
%analyzed in \cite{Ord2}, with
%internal function calls not responding, like external ones in the
%interactive case,
%and with a term-based query causality relation for all function
%symbols, internal and external.
%Queries to the environment or to the ``state manager'' of Section~\ref{sec:model} act the same.
%The difference is that critical terms determine updates to
%internal, not external, locations.}

Since terms might have no value in a given state, the
agreement relation between states must include this possibility.
So, we have
$X =_T Y$, for a set of critical terms $T$, if $\val t X = \val t Y$ for all $t\in T$,
including the case where one of $\val t X$ and $\val t Y$ is $\hole$, in which case both must be.

Imagine that accessing $f(a)$ hangs and the algorithm is simply
$\If x \neq a \Then y := f(x)$.
In contrast to the emulating ASM of Theorem~\ref{thm},
the ASM of the original study hangs even when the algorithm does not,
since it would include the statement
$\If x = a \wedge f(x) =a \Then [\,]$,
which looks at $f(a)$ unnecessarily.
%\begin{ignore}
%For example, that ASM for our sorting example would include the statement in Example~\ref{ex7}.  So, if some $F(i)$ equals the number that happens to be in the out-of-range location $F(n)$, then the emulating ASM will hang on the test $F(i)>F(j)$, when $j=n$, despite the fact that the algorithm itself never accesses $F(n)$.
%\end{ignore}

\begin{example}
Back to sorting.
Suppose now that the elements of the input ``array'' $F$ are computable reals,
represented as programs in some fashion.
And suppose, realistically, that equality is defined for all integers, but not for all reals.
In particular, the inequality relation $>$ (on the representations) gives no result, in general,
when both
arguments represent the same real number, but otherwise behaves as expected.
Then the algorithm in Figure~\ref{fig} sorts $F$ properly when
the $F(i)$ are distinct.
But when $F(i)$ and $F(j)$ represent the same number for two distinct elements in the initial $F$, the
algorithm will not return a result, since at some point it will question whether $F(i) > F(j)$
and receive no answer from the state.
\qed\end{example}

There are idealized models of computation with reals,
such as the BSS model \cite{BSS},
for which real equality is made available.
Still, one may want for division by zero to hang, and for
tests to return true or false, unless a subterm
involves division by zero, in which event the test should also hang.

In practice, one often uses a na{\"\i}ve floating-point approximation to reals,
in which case the results of comparisons are well defined, but may not be very meaningful.
Better, one can deterministically approximate reals by rationally-bounded intervals, for which
arithmetic operations are well-defined, but equality and inequality comparisons
can only provide the ``right'' answer when (non-point) intervals do not overlap.

All the above examples fit perfectly within the framework developed here.

\section{Case Selection}\label{sec:case}

If an algorithm requires information from a state to decide how to proceed, it needs to query the state. The only means it has for this purpose is to pose a question in the form of a term.
The answer to the query must be explicit in the state,
without need for recourse to any algorithm.
Similarly, the algorithm needs to be able to understand answers without further processing, since any such processing ought to be part of the algorithm itself.

We have been assuming,
without loss of generality, that all updates can be postponed until all questions have been asked and answered.
(If not, we need to worry whether subsequent queries use old or new values.)
Furthermore, the question period must eventually end for there to be a next state.

Until now, we have also assumed that all such communication is phrased as Boolean queries,
always resulting in a response of \True\ or \False.%
\footnote{Actually, we treated all values other than \True\ as though they were \False.}
More generally, however, we may presume that there is a fixed, finite set $\K$ of distinct ``distinguished'' values.
Clearly, there need to be at least two elements in $\K$, so that there is more
than one possible outcome of a test, and clearly $\K$ must be finite, or else there would be no
possibility of immediate understanding.
Typically, $\K$ would include, at a minimum, the constants \True, \False, and \Undef.%
\footnote{These are what Kleene \cite{Kleene} suggested in his famous 3-valued logic.}
It is most sensible for each element of $\K$ to have its own self-denoting immutable constant (that is, static nullary) symbol.
So let $|\K|\geq 2$ and let $K$ be symbols carrying those values.

Queried regarding a term $t$, a state $X$ can evaluate $\val t X$ and, if the result is
one of the distinguished values in $\K$, respond with
the symbol $\kappa\in K$ corresponding to that value $\val t X$.
To account for distinguished non-Boolean values, the Discrimination clause of
the Exact Exploration Postulate should be modified
to read as follows:
\begin{itemize}
\item {\bf Discrimination.} \it
For each state $X\in\St$ there is a partial order $\prec_X$ of $\Gamma(X)$ such that for every other state $Y\in\St$, and for every $t\in \Gamma(X)\setminus\Gamma(Y)$,
if any, there is a \emph{discrimination} term $s \in \Gamma(X)$ such that $s \prec_X t$ and $\val s X=\val \kappa X$, $\val s Y=\val \lambda Y$, for distinct nullaries $\kappa,\lambda \in K$.
\end{itemize}

To reflect the more complicated case analysis, with a richer than Boolean class of distinguished values
governing intra-step behavior, an ASM can use compound \textbf{case} statements.%
\footnote{Case statements were introduced in the Pascal programming language.
They resemble the switch statement of the C family of languages.
The syntax we chose is akin to that of SQL.}
The form of the statement is
\[
\Case q_1,\dots,q_n \Of
\begin{cases}
\When a_{11},\dots,a_{1n} \Then& S_1 \;;\\[0pt]
\When a_{21},\dots,a_{2n} \Then& S_2  \;;\\
 \qquad\vdots \\
\When a_{m1},\dots,a_{mn} \Then& S_m
\end{cases}
\]
Each of the $q_j$ is a term;
each of the $a_{ij}$, etc.\ is a nullary element of $K$.
The cases $\bar a_i$ need not be mutually exclusive.%
\footnote{The syntax could be sugared to also allow wildcards or a ``catch-all'' case.}
A statement $S_j$ can be an update
or another case statement, or a parallel mix of updates and/or cases.

The \textbf{case} statement supplants the \textbf{if} statement we used earlier.%
\footnote{It also subsumes parallel composition, leaving out query terms.}
It is not, of course, strictly necessary, because one could use parallel conditionals of the following form, instead:
%\nb{can omit one line of cases}
\begin{equation}\label{if}
\begin{array}{lll}
[ &
\If q_1=a_{11} \wedge \dots \wedge q_n=a_{1n} \Then S_1 &  \|  \\
& \If q_1=a_{21} \wedge\dots \wedge q_n=a_{2n} \Then S_2 & \|  \\
%\| ~\cdots~\| \\
& \qquad \vdots & \\[-1mm]
& \If q_1=a_{m1} \wedge\dots \wedge q_n=a_{mn} \Then S_m &]
\end{array}
\end{equation}

Operationally, the following transpires:
The state is ``queried'' for the values of all the terms $q_1,\dots,q_n$.
The locations corresponding to these terms $q_i$ are, indeed, explored.
If the results agree with one or more of the listed cases of values $a_{j1},\dots,a_{jn}$,
then each such $S_j$ is evaluated and whatever
needs to be explored in the evaluation process is also explored.
Other components $S_k$, those for which the $q_i$ do not match the $a_{ik}$, are not explored.
If the results agree with none of the cases, then the statement does nothing other than
locate and explore the queried locations.
When, for some state $X$, one of the query terms $q_i$ does not yield a distinguished value in $\K$,
the case statement fails and
the whole algorithm comes to a standstill;
then we will have
$\Delta(X)=\hole$.

The semantics of ASMs is extended with the following clauses:
\begin{equation}\label{eq:ext}
\renewcommand\arraystretch{1.2} %\setlength{\arraycolsep}{4pt}
\begin{array}{lll}
\Delta^+_{\Case \bar q \Of \overline W}(X) &=& \bigcup\nolimits_j D^{\bar q}_{W_j}(X)\\
D^{\bar q}_{\When \bar a \Then P}(X) &=&
\begin{cases}
\Delta^+_P(X) & \mbox{if $X\models \bar q = \bar a$}\\[1ex]
\{\hole\} & \mbox{if $\bar a\notin \K^n$}\\
\emptyset & \mbox{otherwise}
\end{cases}
\\
\Gamma_{\Case \bar q \Of \overline W}(X) &=& \{q_1,\dots,q_n\}\cup \bigcup\nolimits_j G^{\bar q}_{W_j}(X)\\
G^{\bar q}_{\When \bar a \Then P}(X) &=&
\begin{cases}
\Gamma_P(X) & \mbox{if $X\models \bar q = \bar a$}\\[1mm]
\emptyset & \mbox{otherwise}\;.
\end{cases}
\end{array}
\end{equation}
Actually, the constant terms in $K$ are also explored, but we may just as well assume that they are all
explored in all states, and not bother include them explicitly in $\Gamma$
(much as we did not explicitly include \True{} and \False{} until now).

%As can be seen from the above formulation, % Eq.~(\ref{if}),
One need only presume that states can recognize equality with the distinguished values in $\K$.
Equality between other values can be partial, as in Section~\ref{sec:eq}.

Now, we modify the definition of the update set for ASMs to incorporate case statements
and black holes as follows:
\[
\Delta(X) = 
\begin{cases}
\hole & \mbox{if $\hole\in\Delta^+(X)$}\\
\bot & \mbox{if $\Delta^+(X)\in\{\emptyset,\bot\}$}\\
\Delta^+(X) \setminus \Delta^0(X) & \mbox{otherwise}\;.
\end{cases}
\]

The compound case condition makes it easier to model the possibility of an algorithm posing questions for which it may not end up caring about the answer, nor caring about the order in which that batch of questions is answered.

%\begin{remark}
Redundancies can be omitted.
In other words, if any path asks the same query term twice, the inner
one may be removed.
By Bounded Exploration, we know that under all circumstances
only a finite number of different questions can be asked.
So for a given set of critical terms, only a finite number
of irredundant case statements are possible.
%\end{remark}

It is straightforward to reprove Theorem~\ref{thm}
for algorithms with richer-than-Boolean queries.
Instead of creating parallel conditionals when $V$
is not agreeable, create a case statement.
Using the case statement makes it an easy matter to build
a unique, ``normal-form'' emulating ASM for any exacting algorithm.
%\nb{uniqueness}

\section{Three Variations}\label{sec:vary}

\subsection{Access}\label{sec:access}

Locations are used in tests, to extract contents, and for addressing.
Consider ASM programs.
\begin{enumerate}
\item We have seen how critical terms are used in the conditional tests of \textbf{if} statements
and in the queries of the \textbf{case} statement.  The values in the indicated locations determine what else transpires.
\item We have also seen that the term on the right-hand side of an assignment is one
of the critical terms (Lemma \ref{lem}).
The content of such a location is copied into another location as part
of an update.
\item Lastly, critical terms are used indirectly to determine locations needed for tests or updates.
(Depending on the internal workings of states, that determination may involve equality checks.)
\end{enumerate}

In the statement $\If p(x) \Then f(g(x)) := h(f(x))$, the critical term $p(x)$ points to a Boolean
location of the first variety,
$h(f(x))$ is of the second, and $x$, $f(x)$, and $g(x)$ serve the third purpose.

When a location is used for copying, the state manager (see Section~\ref{sec:model}) need not understand anything about it.
It simply needs to copy the contents, as is, to another location.
On the other hand, locations whose values are used as tests
must be understood.

So, one might want to partition $\Gamma(X)$ into three:
$\Gamma^D(X)$ for the discriminating terms used
in conditional and case statements;
$\Gamma^C(X)$ for obtaining the contents of locations indicated by right-hand sides of assignments;
$\Gamma^A(X)$ for addressing locations.
Looking back at the computation of $\Gamma$ for ASMs in Eqs.~(\ref{eq:gamma},\ref{eq:ext}),
the allocation of terms to the parts of $\Gamma$ is as follows:
\begin{itemize}
\item
in the assignment case, the $s_i$ go into $\Gamma^A$,
while $t$ goes into $\Gamma^C$;
\item
in the conditional case, the condition $C$ goes into $\Gamma^C$;
\item
the same goes for case-statement queries $q_i$;
\item
subterms of everything in any part of $\Gamma$ also go into $\Gamma^A$.
\end{itemize}

One can ascertain that the construction of the emulating ASM respects this partition of
uses of explore terms.
%%\begin{ignore}
%%Note also that some aspects of intra-step behavior of an algorithm, like the number of times a location is
%%accessed, are not taken into consideration by the emulation given here.
%%\end{ignore}
%\begin{comment}
%Not to be added to the paper, but just to be noted for our own use:  We don't attempt here to axiomatize this partition of $\Gamma(X)$.  It's discussed only in the ASM context.
%\end{comment}

\subsection{Clash}\label{sec:clash}

Clashes are used by ASMs to model failure.
When an attempt is made to assign different values to the same location,
the standard semantics of ASMs says that the update fails and there is no next state.
Until now, we have not considered such ASMs.

To model such behavior, we need to emulate all proposed updates, even
when there is no next state.
To that end, $\Delta$ in the Determination clause of Postulate~\ref{EE} should be
changed to $\Delta^+$.
Furthermore, there need to be some behind-the-scenes tests.
For any state $X$, with proposed updates $\Delta^+(X)$ (Eq.~\ref{eq:prop}),
we might add the following two sets of equality-terms to $\Gamma(X)$:
\[
\begin{array}{c}
\{ u_i=v_i \st f(\bar u):=s, f(\bar v):= t \in \Delta^+(X),~u_i\not\equiv v_i ~s\not\equiv t \}
\\[4pt]
\{ s=t \st f(\bar u):=s, f(\bar v):= t \in \Delta^+(X),~\val{\bar u=\bar v}{X},~s\not\equiv t \} \;,
\end{array}
\]
where $\equiv$ denotes syntactic identity of terms,
to reflect the implicit tests for clashes.
The ordering $\prec_X$ should place (the new terms in) these after the part of $\Gamma(X)$
obtained from the tests,
both followed by the terms contributed by the assignments themselves.
%\nb{an example if there's space}

\subsection{Bounding Explore Sets}

An algorithm, when presented with a state $X$, can only perform a finite amount of work on $X$
in one step.
Work, here, includes the exploration of locations, and perhaps updating some of them.
As explained above, the only means by which an algorithm can identify specific locations within a state
is via terms.
So, we clearly must have $|\Gamma(X)|<\infty$ for all states $X$, as expressed in Postulate~\ref{EE}.

Just because each explore set $\Gamma(X)$ is finite does not, of course,
mean that the sum total of explore terms is finite, since different terms
can appear in each set.
Because an algorithm must be finitely describable,
Postulate~\ref{EE} includes the Limitation clause.
As it turns, out, however, that clause can be stated in a weaker form.

\begin{theorem}
The following formulations of the Limitation clause of Postulate~\ref{EE} are equivalent:
\begin{enumerate}
\item There are only finitely many explore terms: $\left|\bigcup_{X\in\St}\Gamma(X)\right|<\infty$.
\item There is a uniform bound on explore sets: for some bound $N$, $|\Gamma(X)|<N$ for all states $X\in\St$.
\item There are no infinite sequences $X_1, X_2, \ldots$ of states and $s_1 , s_2 , \ldots$ of terms such that for all $i$, we have $s_1 \prec_{X_i} s_2 \prec_{X_i} \cdots \prec_{X_i} s_{i}$, 
where
$s_i$ discriminates $X_i$ from all $X_j$, $j>i$.
\end{enumerate}
\end{theorem}

\begin{pf}
Clearly (a) implies (b).  And (b) implies (c), because the chain of distinct $s_i$'s in (c) cannot comprise more than the $N$ terms of (b).

Every state $X$ corresponds to a path in the tree of Figure~\ref{tree},
with $\Gamma(X)$ consisting of all terms in the path's nodes.
Thus, by Lemma~\ref{lem:less}, terms get bigger along paths, so
(c) implies that \emph{all} paths are finite
(whether conditions are rich or Boolean,
and without recourse to overall finiteness, as needed for Theorem~\ref{thm}).
By K\"onig's Lemma, (a) follows.%
\footnote{Cf.\ the related argument in \cite[Theorem~A.4]{Gurevich00}, which
invokes the compactness of Cantor spaces rather than K\"onig's (weak) Lemma.}
\end{pf}

\section{Discussion}

We have shown that every classical algorithm, which satisfies natural postulates,
can be step-by-step emulated by an abstract state machine that does not
attempt to apply equality or functions to more values than does the algorithm.
This significantly strengthens the thesis, propounded in \cite{Gurevich00},
that abstract state machines faithfully model
any and all sequential algorithms.

The easing of the requirements on fully defined equality and other functions
lends strong support to the contention---put forth in \cite{CTT,CT}---%
that the Church-Turing Thesis is provably true from first principles.
In addition to the Sequential Postulates,
the arguments for Church's Thesis require that
initial states contain only free constructors and functions that
can be programmed from constructors.
Our refinement of the ASM Theorem strengthens those results
by showing that the simulation of an algorithm, having no (unprogrammable) oracles, by
an effective abstract state machine
need not involve any operations not available to the original algorithm.
It also follows from this work that there is no harm in
incorporating partial operations in the initial states of effective algorithms,
as long as they too can be computed effectively.
Even with this relaxation of the limitations on initial states,
it remains provable that no super-recursive function can be computed algorithmically.

\subsection*{Acknowledgements}
We thank Olivier Bournez for his comments.

\newpage
\addcontentsline{toc}{section}{References}
%\bibliographystyle{plain}
%\bibliography{Part}

\end{document}